\documentclass[11pt]{article}
\usepackage[margin=0.85in]{geometry}
\usepackage[T1]{fontenc}
\usepackage[utf8]{inputenc}
\usepackage{lmodern}
\usepackage{microtype}
\usepackage{amsmath,amssymb,bm}
\usepackage{graphicx}
\usepackage{float}
\usepackage{booktabs}
\usepackage{array}
\usepackage{tabularx}
\usepackage{enumitem}
\usepackage{xcolor}
\usepackage{caption}
\usepackage[numbers,sort&compress]{natbib}
\usepackage[hidelinks]{hyperref}
\usepackage{url}

\newcolumntype{Y}{>{\raggedright\arraybackslash}X}
\setlist{nosep,leftmargin=*}
\hypersetup{
  pdftitle={EAF-TTS: Rethinking Codec-Latent Coordinates for Flow TTS},
  pdfauthor={Yumiao Li, Haoxiang Jie, Xinbei Li, Yaoyuan Yan, Yue Qu},
  pdfsubject={Text-to-speech, flow matching, codec latents, ZCA whitening, few-call inference},
  pdfkeywords={text-to-speech, flow matching, codec-latent TTS, latent interface, latent whitening, ZCA, proxy naturalness, sampling efficiency}
}

\title{EAF-TTS: Rethinking Codec-Latent Coordinates for Flow TTS}
\author{%
Yumiao Li$^{1}$ \quad Haoxiang Jie$^{1}$ \quad Xinbei Li$^{1}$ \quad Yaoyuan Yan$^{1}$ \quad Yue Qu$^{2}$\\
\small $^{1}$AI Lab, Country Garden Services Group\\
\small $^{2}$Tongji University\\
\small \texttt{liyumiao02@bgyfw.com}, \texttt{jiehaoxiang@bgyfw.com}
}
\date{}

\begin{document}
\maketitle

\begin{center}
\small\emph{Preprint. UTMOS is treated throughout as an automatic proxy for naturalness, not as human MOS. The repeatedly accessed Test399 split is used only for diagnosis and is not presented as a blind test set.}
\end{center}

\begin{abstract}

Latent text-to-speech (TTS) systems generate codec latents decoded by a frozen decoder, but good reconstruction does not ensure geometry suitable for Flow generation from isotropic Gaussian noise. We present Embedded Acoustic Flow TTS (EAF-TTS), an acoustic-latent adaptation of ELF with phone- and duration-conditioned generation, shared-network terminal correction, and frozen WaveVAE decoding. EAF-TTS supports a fixed ZCA interface estimated from training data and adding no learned parameters. The codec, conditioning modules, Flow backbone, and 20-call inference are shared. The intervention is the full ZCA coordinate protocol---training and sampling coordinates, induced loss metric, and terminal perturbation---not an inference-only transform. In single-speaker LJSpeech, the 64-D training-latent covariance has condition number 1355.44. On text-disjoint Val399 strict-GT with fixed E100 raw checkpoints across three matched seeds, ZCA improves UTMOS22 Strong---an automatic naturalness proxy, not human MOS---by $+0.405\pm0.027$ over Scalar normalization (mean $\pm$ seed SD of per-seed paired differences). Absolute scores remain below codec-reference levels. Without retraining, ZCA-5 retains $+0.375\pm0.027$ UTMOS over Scalar-20 while reducing generation-path RTF by $3.34\times$. This three-seed analysis shows no statistically detectable WER or exact-match degradation, but does not establish equivalence. ZCA-3 gives $5.03\times$ with a $+0.165$ UTMOS advantage. At matched reduced call counts, ZCA's advantage shrinks, indicating full-step headroom rather than better low-step integration. A secondary text-only selected-checkpoint result is directionally consistent but confounded by checkpoint type. We claim only proxy-naturalness and measured quality--latency effects within this protocol; human listening, blind text, and broader validation remain necessary.

\end{abstract}

\section{Introduction}

Latent generative models compress high-rate waveforms into low-rate continuous representations and learn a conditional distribution over them, shortening the generated sequence and leaving fine waveform reconstruction to a separately trained neural audio codec. Codec-latent generation has proved effective in NaturalSpeech 2 \cite{shen2023naturalspeech2}. Separately, Matcha-TTS, Voicebox, and F5-TTS have demonstrated efficient and scalable Flow Matching for speech synthesis \cite{mehta2023matcha,le2023voicebox,chen2024f5tts}. Building on Embedded Language Flows (ELF) \cite{hu2026elf}, we introduce Embedded Acoustic Flow TTS (EAF-TTS), an acoustic-latent adaptation with two shared-network modes: mode 0 performs phone- and duration-conditioned Flow generation, and mode 1 performs one terminal clean-latent correction before frozen WaveVAE decoding. EAF-TTS is the platform for the controlled coordinate study below.

This modular design introduces an interface problem. A codec is trained for reconstruction, not for isotropic Gaussian priors, straight probability paths, or Euclidean Flow Matching objectives. Good reconstruction establishes $D(E(s))\approx s$ for waveform $s$, where $E$ and $D$ denote the codec encoder and decoder, but it does not establish that $E(s)$ is well conditioned for conditional Flow generation. The codec need not itself be trained for these generative objectives; the question is whether a post-hoc coordinate protocol can compensate. On the repeatedly accessed Test399 diagnostic set, original audio and frozen-WaveVAE reconstructions obtain UTMOS22 Strong scores of 4.372 and 4.259 and ASR WERs of 2.885\% and 2.856\%, respectively. In the seed-42 fixed-E100-raw strict-GT diagnostic run, the Scalar Flow obtains 3.162\% WER but only 2.785 UTMOS22 Strong. This contrast provides diagnostic evidence consistent with localizing the observed proxy-naturalness gap to conditional latent generation. These numbers motivate the question; they are non-blind and are not used to estimate the ZCA effect, which is evaluated on Val399 with fixed E100 raw checkpoints.

The covariance of the 64-dimensional LJSpeech training latents motivates the intervention: it has condition number 1355.44, with eigenvalue range $[0.0061128,8.28546]$. Scalar normalization corrects only global mean and scale, leaving channel correlation and principal-direction anisotropy intact. ZCA is a canonical symmetric whitening choice \cite{kessy2018optimal}, but the frozen decoder's directional sensitivity need not align with data covariance. We therefore treat ZCA as a complete coordinate protocol and audit its decoder-domain and optimization effects rather than assuming that a lower condition number is sufficient.

We ask a deliberately narrow question:

\begin{quote}\itshape
Under a frozen codec and a shared generative architecture, does the full ZCA coordinate protocol, estimated only from training data, reproducibly improve UTMOS22 Strong proxy naturalness across three matched seeds on text-disjoint Val399 under strict ground-truth durations?
\end{quote}

Scalar and ZCA systems use the same WaveVAE, text and phone front ends, duration model, EAF-TTS backbone, training budget, matched random seeds, and, for the primary comparison, the same inference settings. The intervention is the full ZCA coordinate protocol: a fixed train-data-estimated transform with no learned parameters, its induced Euclidean loss metric when expressed in codec coordinates, and the coordinate-space perturbation used to train mode 1. The transform is invertible up to audited numerical precision. We interpret the results as the aggregate effect of this protocol, not as evidence that reducing the condition number alone is causal. The primary analysis compares fixed E100 non-EMA weights across three matched seeds. A zero-retraining call-count sweep is a deployment audit rather than a pure coordinate comparison. A secondary text-only comparison is also non-causal and additionally confounded by its use of Scalar EMA and ZCA raw checkpoints.

We do not claim novelty in Flow Matching, codec latents, explicit durations, or whitening/ZCA. EAF-TTS is a specific acoustic-latent adaptation of ELF; the primary empirical contribution is the controlled treatment of codec-latent second-order geometry within this system. Our contributions are:

\begin{enumerate}
\item \textbf{EAF-TTS.} We introduce an acoustic-latent adaptation of ELF that combines phone- and duration-conditioned continuous generation, shared-network terminal correction, and frozen WaveVAE decoding.
\item \textbf{Diagnostic localization.} Non-blind reconstruction diagnostics provide evidence consistent with localizing the observed proxy-naturalness gap to conditional latent generation, rather than to frozen-codec reconstruction, under the studied protocol.
\item \textbf{Controlled full-step ZCA effect.} We evaluate the full ZCA coordinate protocol with latent round-trip and waveform-reconstruction audits. On Val399 strict-GT with fixed E100 raw checkpoints, it improves UTMOS22 Strong by $+0.405\pm0.027$ over Scalar across three matched seeds, while WER, exact-match rate, and DTW-MFCC do not improve consistently.
\item \textbf{Few-call and boundary evidence.} ZCA-5 retains $+0.375\pm0.027$ UTMOS over Scalar-20 while reducing measured strict-GT generation-path RTF by $3.34\times$. The fixed-three-seed analysis finds no statistically detectable WER or exact-match degradation, but does not establish equivalence. At every matched reduced call count, ZCA's advantage over Scalar is smaller than at 20 calls, indicating usable full-step headroom rather than better low-step integration. The confounded text-only comparison and negative decoder and gradient audits are retained as secondary boundary evidence.
\end{enumerate}

\section{Related Work}

\subsection{Flow Matching for acoustic latent generation}

Flow Matching trains continuous normalizing flows by regressing a vector field along a prescribed probability path \cite{lipman2022flow}. Matcha-TTS applies optimal-transport conditional Flow Matching to mel-spectrogram generation in a non-autoregressive TTS architecture \cite{mehta2023matcha}; we treat explicit duration conditioning as a fixed system component rather than as a contribution. Voicebox and F5-TTS demonstrate the scalability of Flow Matching for large speech-generation systems \cite{le2023voicebox,chen2024f5tts}.

We introduce EAF-TTS as our acoustic-latent adaptation of Embedded Language Flows (ELF) \cite{hu2026elf}. ELF performs most generation in continuous token-embedding space and switches to a shared-weight decoding mode only at the endpoint. EAF-TTS transfers this pattern to continuous acoustic latents with two shared-network modes: mode 0 performs phone- and duration-conditioned Flow generation, whereas mode 1 makes one terminal clean-latent correction before frozen WaveVAE decoding. To the best of our knowledge, this specific ELF-to-TTS adaptation has not previously been described. EAF-TTS does not introduce a new general Flow objective; it instantiates ELF's shared generation/terminal-correction pattern in an acoustic-latent TTS system.

\subsection{Codec-latent generation}

Neural audio codecs such as SoundStream show that low-rate representations can preserve high-quality waveform information \cite{zeghidour2022soundstream}. NaturalSpeech 2 performs speech and singing generation with codec-latent diffusion \cite{shen2023naturalspeech2}. SupertonicTTS combines continuous speech-autoencoder latents, Flow-Matching text-to-latent generation, and an utterance-level duration predictor \cite{kim2025supertonic}. Beyond neural audio codecs, LAFMA studies Flow Matching on continuous mel-spectrogram VAE latents for text-to-audio generation \cite{guan2024lafma}, while Flow Matching in Latent Space studies Flow generation in continuous image-autoencoder representations \cite{dao2023latent}. EAF-TTS differs by adapting ELF's shared-network generation/terminal-correction pattern to phone- and duration-conditioned WaveVAE latents. The present experiment then freezes that architecture and isolates a global second-order (covariance) coordinate intervention.

\subsection{Whitening and preconditioning}

Whitening centers data and applies an inverse covariance square root. Whitening matrices are not unique. For a positive-definite covariance, ZCA uses its unique symmetric positive-definite inverse square root and, among whitening transforms, minimizes the expected squared distance between centered original and whitened variables \cite{kessy2018optimal}. Decorrelated Batch Normalization and VAE whitening show that whitening can alter optimization and representation factorization \cite{huang2018decorrelated,hahn2018disentangling}. Preconditioned diffusion sampling uses matrix preconditioning to mitigate ill-conditioned sampling geometry \cite{ma2022preconditioned}, whereas conditionally whitened generative models incorporate conditional means and covariances into diffusion and Flow Matching \cite{yang2026conditionally}.

These results do not guarantee that whitening codec latents will improve TTS in general. A frozen decoder induces a decoder-domain geometry that need not match data covariance, and a coordinate transform changes both squared losses and decoder-domain gradients. We therefore pair positive generation results with audits that constrain a simple ``lower condition number is sufficient'' explanation.

\section{Method}

\subsection{Embedded Acoustic Flow TTS}

The frozen WaveVAE encodes waveform $s$ as

\[
z=E(s),\qquad z\in\mathbb{R}^{T\times64},
\]

where $T$ is the acoustic frame count shared by the codec latent and the frame-level condition.

Text is normalized and converted to ARPA phone symbols with a lexicon and a grapheme-to-phoneme (G2P) module. Following the explicit length-expansion paradigm of FastSpeech-style systems \cite{ren2020fastspeech2}, in which phone durations are predicted before acoustic generation, a frozen duration model predicts total duration, phone durations, and pauses. A length regulator expands phone representations into frame-level condition $C\in\mathbb{R}^{T\times h}$. The acoustic sequence length and phone timestamps are therefore determined before ODE integration. Figure 1 summarizes the full pipeline and the placement of the coordinate interfaces.

This adaptation has three practical properties. First, generation operates on the codec's low-rate latent sequence rather than at waveform rate, leaving fine waveform reconstruction to the frozen decoder. Second, explicit durations determine the latent length and phone timestamps before integration. Third, mode 1 reuses the mode 0 backbone for terminal correction and therefore does not introduce a separate large refiner. These are architectural properties; the present study does not isolate their individual causal effects.

Modes 0 and 1 share the latent input and output projections, positional representation, phone encoder, frame and speaker conditioning, DiT blocks, and output head. A learned two-entry mode embedding, added to the time embedding, selects the behavior. Mode 1's latent, mel, and STFT losses update these shared EAF-TTS parameters. WaveVAE parameters remain frozen, while gradients propagate through its decoder from the mel and STFT objectives to the predicted latent.

\begin{figure}[H]
\centering
\includegraphics[width=\textwidth]{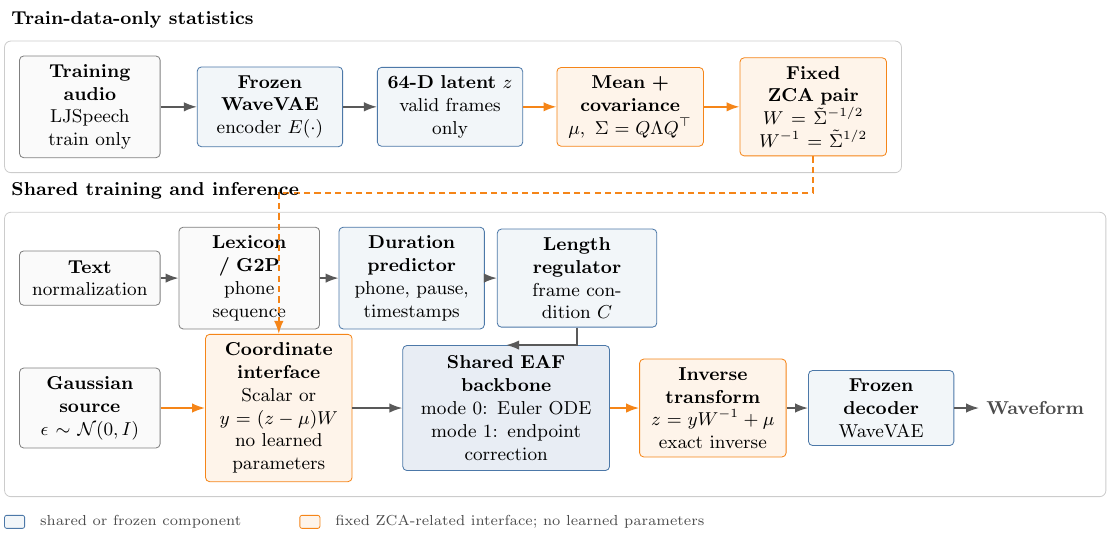}
\caption{ZCA EAF-TTS with train-data-only statistics. Orange modules are fixed ZCA-related coordinate interfaces with no learned parameters; blue modules are shared or frozen components. Scalar and ZCA systems share the text front end, duration model, EAF backbone, and frozen WaveVAE. The fixed ZCA statistics are estimated from training data only.}
\end{figure}

For a target latent in the chosen coordinate system, denoted $x\in\mathbb{R}^{T\times64}$ (see Section 3.2 for the Scalar and ZCA choices), a Gaussian source $\epsilon\in\mathbb{R}^{T\times64}$ with $\epsilon\sim\mathcal N(0,I)$, and time $t\in(0.01,0.99)$ drawn from the logit-normal schedule shared by the Scalar and ZCA mode 0 systems (Section 4.3), the training path is

\[
x_t=t x+(1-t)\epsilon.
\]

The shared mode 0 network predicts a clean endpoint $\hat x_\theta$. Training minimizes a time-weighted endpoint-prediction loss, averaged over valid frames:

\[
\mathcal L_{\mathrm{FM}}
=\mathbb E\!\left[
\frac{\lVert\hat x_\theta(x_t,t,C)-x\rVert_2^2}
{\left(1-\min(t,0.95)+10^{-5}\right)^2}
\right].
\]

The time cap limits the maximum near-terminal weight to approximately $400$, even when the sampled $t$ exceeds $0.95$. The velocity is constructed only for numerical integration at inference:

\[
v_\theta(x_t,t)=\frac{\hat x_\theta-x_t}{1-t}.
\]

Primary full-step inference uses 20 model calls: 19 mode 0 Euler updates from $t=0$ to $t_{\mathrm{end}}=0.95$ with $\Delta t=t_{\mathrm{end}}/19=0.05$, followed by one mode 1 clean-latent correction at $t=1$. The endpoint $t_{\mathrm{end}}=0.95$ is a stability choice rather than a theoretical requirement. Because velocity reconstruction divides by $1-t$, a fixed clean-endpoint prediction error is amplified by approximately $20\times$ at $t=0.95$ and $100\times$ at $t=0.99$, while the expression is undefined at $t=1$. An exact predictor would cancel the numerator in the limit, but finite prediction errors can destabilize near-terminal integration under explicit integrators such as Euler. We therefore stop mode 0 integration at $t_{\mathrm{end}}$ and use mode 1 as a learned terminal correction that directly predicts the clean latent without constructing a velocity. Section 4.4 evaluates fewer Euler updates at the same endpoint without retraining.

On the ideal interpolation path, $x_{t_{\mathrm{end}}}=t_{\mathrm{end}}x+(1-t_{\mathrm{end}})\epsilon$ retains only a $0.05$ source-noise coefficient and is therefore close to the clean endpoint. Mode 1 is trained in each system's active Flow coordinate on clean latents with small Gaussian perturbations applied to a subset of frames. The numeric perturbation scale is 0.1 for both Scalar and ZCA, but the same coordinate-space magnitude does not imply the same codec-space perturbation after their different inverse transforms. The primary comparison therefore includes this coordinate-dependent mode 1 training effect. The generated state at $t_{\mathrm{end}}$ is intended to lie in the near-clean input regime, although the learned ODE state need not follow the ideal path exactly and the training and inference input distributions are not assumed to match. Mode 1 uses latent, mel, and STFT auxiliary losses with the shared weights reported in Section 4.3 and requires no clean target at inference.

\subsection{Scalar and ZCA coordinates}

The Scalar baseline applies a single global scalar mean $\mu_s\in\mathbb R$ and scalar standard deviation $\sigma_s\in\mathbb R$, estimated jointly over all valid training frames and all 64 channels:

\[
x_s=\frac{z-\mu_s}{\sigma_s}.
\]

Let $\boldsymbol\Sigma$ denote the channel covariance of the original codec latents, estimated over valid training frames. Scalar normalization changes it only to $\boldsymbol\Sigma/\sigma_s^2$, so the covariance condition number remains $1355.44$ and the eigenvectors are unchanged; channel correlations and principal-direction anisotropy are therefore preserved.

ZCA instead uses the same $N$ valid training frames, with each row vector $z_n\in\mathbb R^{1\times64}$, to estimate the channel mean and population covariance (normalized by $N$, not $N-1$):

\[
\boldsymbol\mu=\frac{1}{N}\sum_{n=1}^{N}z_n,\qquad
\boldsymbol\Sigma=\frac{1}{N}\sum_{n=1}^{N}
(z_n-\boldsymbol\mu)^\top(z_n-\boldsymbol\mu),
\quad \boldsymbol\Sigma\in\mathbb R^{64\times64}.
\]

Using the row-vector convention and the eigendecomposition $\boldsymbol\Sigma=Q\Lambda Q^\top$, we stabilize the eigenvalues as $\tilde\lambda_i=\max(\lambda_i,10^{-5}\lambda_{\max},10^{-8})$, where $10^{-5}\lambda_{\max}$ is a relative floor and $10^{-8}$ is an absolute floor, and write $\tilde\Lambda=\operatorname{diag}(\tilde\lambda_i)$. For LJSpeech, the raw eigenvalue range is $[0.0061128,8.28546]$, giving $8.28546/0.0061128\approx1355.4$. The relative floor is $8.28546\times10^{-5}$ and the absolute floor is $10^{-8}$; both are below $\lambda_{\min}$, and no eigenvalue is clamped. The stabilization is therefore a numerical safety net rather than a factor in the reported LJSpeech results. The forward and inverse transforms are

\[
y=(z-\boldsymbol\mu)Q\tilde\Lambda^{-1/2}Q^\top,
\qquad
z=yQ\tilde\Lambda^{1/2}Q^\top+\boldsymbol\mu.
\]

Thus, the target coordinate $x$ in Section 3.1 is $x_s$ for the Scalar system and $y$ for the ZCA system. The ZCA Flow is trained and sampled in $y$; $z$ is recovered before WaveVAE decoding. The transform changes neither latent dimension nor frame rate and adds no learned parameters, making it the fixed coordinate interface shown in Figure 1. Its statistics are estimated once from valid training frames and then frozen. Round-trip and codec-invariance audits pass the pre-specified implementation thresholds: latent round-trip maximum absolute error below $10^{-4}$, waveform mean absolute error below $10^{-4}$, and waveform maximum absolute error below $2\times10^{-3}$.

ZCA is not optimization-neutral. Let $\tilde{\boldsymbol\Sigma}=Q\tilde\Lambda Q^\top$ denote the stabilized covariance used by the transform, whereas $\boldsymbol\Sigma$ is the raw training covariance reported above. Let $\hat z=\hat yQ\tilde\Lambda^{1/2}Q^\top+\boldsymbol\mu$ be a prediction mapped back to codec coordinates. Euclidean error in whitened coordinates is a stabilized Mahalanobis error in codec coordinates:

\[
\|\hat y-y\|_F^2
=\operatorname{tr}\!\left[(\hat z-z)\tilde{\boldsymbol\Sigma}^{-1}(\hat z-z)^\top\right].
\]

Here the trace sums the per-frame Mahalanobis errors over the $T$ acoustic frames. Thus, training in $y$ changes the effective loss metric in codec coordinates, not only the coordinate system. We therefore interpret our results as the joint effect of the full ZCA coordinates and their induced optimization metric, not as causal evidence that reducing the condition number alone is responsible.

\subsection{Inference conditions}

We distinguish three inference paths:

\begin{enumerate}
\item \textbf{Strict-GT:} MFA phones and ground-truth phone durations, isolating acoustic generation.
\item \textbf{Cached predicted duration:} outputs from the frozen duration model, computed once, cached, and reused across systems.
\item \textbf{Text-only:} lexicon/G2P and duration prediction run at inference time from raw text; no MFA alignment, phone cache, or duration cache is read.
\end{enumerate}

The three-seed primary analysis uses Val399 strict-GT with fixed E100 non-EMA (raw) checkpoints; text-only results evaluate the selected deployment systems.

\section{Experimental Setup}

\subsection{Data}

LJSpeech is resampled to 24 kHz and split by normalized-text key:

\begin{table}[H]
\centering
\caption{LJSpeech data splits and their roles in this study.}
\label{tab:data}
\small
\setlength{\tabcolsep}{4pt}
\begin{tabularx}{\textwidth}{lrY}
\toprule
Split & Utterances & Use \\
\midrule
Train & 12,080 & model training and ZCA statistics \\
Validation & 399 & checkpoint selection and primary results \\
Test & 399 & diagnostics only; repeatedly accessed during development and not treated as blind \\
\bottomrule
\end{tabularx}
\end{table}

The splits have no normalized-text overlap. WaveVAE was not trained on validation or test audio. ZCA statistics use 3,686,471 valid Train latent frames and exclude padding. Val399 is text-disjoint but is a development validation set rather than a prospectively untouched blind set; the primary comparison nevertheless fixes E100 raw checkpoints for all six runs instead of selecting a checkpoint separately for each system.

\subsection{Latent statistics}

\begin{table}[H]
\centering
\caption{Train-set WaveVAE latent statistics and phone-conditioned covariance diagnostics.}
\label{tab:latent-stats}
\small
\setlength{\tabcolsep}{4pt}
\begin{tabular}{lr}
\toprule
Statistic & Value \\
\midrule
Latent channels & 64 \\
Minimum / maximum eigenvalue & 0.0061128 / 8.28546 \\
Condition number & 1355.44 \\
Effective rank & 22.18 \\
Participation ratio & 10.35 \\
Within-lexical-phone condition & 524.98 \\
Post-ZCA within-phone condition & 5.06 \\
\bottomrule
\end{tabular}
\end{table}

For normalized eigenvalues $p_i=\lambda_i/\sum_j\lambda_j$, effective rank is $\exp(-\sum_i p_i\log p_i)$ and participation ratio is $(\sum_i\lambda_i)^2/\sum_i\lambda_i^2$. The within-lexical-phone statistic is not a mean of per-phone condition numbers. It first forms the frame-weighted pooled within-phone covariance

\[
\boldsymbol\Sigma_{\mathrm{within}}
=\frac{1}{N_{\mathcal P}}\sum_{p\in\mathcal P}N_p\boldsymbol\Sigma_p,
\]

where $\mathcal P$ contains lexical phones only, $N_p$ is the number of frames assigned to phone $p$, $N_{\mathcal P}=\sum_{p\in\mathcal P}N_p$, and padding, unknown, silence, and short-pause tokens are excluded. The reported condition numbers are computed with a relative eigenvalue floor of $10^{-6}$, before and after applying the global ZCA matrix to $\boldsymbol\Sigma_{\mathrm{within}}$. This is a diagnostic floor, distinct from the $10^{-5}$ relative floor used to construct the global ZCA transform in Section 3.2; the smaller diagnostic value avoids unnecessarily clamping low-variance within-phone directions. No eigenvalue of the pooled lexical within-phone covariance, either before or after global ZCA, falls below this diagnostic floor, so the floor is inactive and the reported conditions reflect the raw pooled covariances. This audit confirms that ZCA improves conditioning beyond global between-phone variation; it does not guarantee improved generation quality.

\subsection{Training and reproducibility}

Scalar and ZCA systems share the frozen WaveVAE, phone vocabulary, text front end, duration model, EAF network, optimizer, schedule, auxiliary losses, and inference settings. We train matched pairs for seeds 42, 2027, and 3407. The primary analysis fixes E100 raw checkpoints. The seed-2027 text-only comparison instead uses checkpoints selected by an existing validation rule that was not publicly preregistered: Scalar E100 EMA versus ZCA E100 raw. It is a deployment comparison, not a pure coordinate estimate.

\begin{table}[H]
\centering
\caption{Shared training and inference configuration for the Scalar and ZCA systems.}
\label{tab:training}
\small
\setlength{\tabcolsep}{4pt}
\begin{tabularx}{\textwidth}{>{\raggedright\arraybackslash}p{0.25\textwidth}Y}
\toprule
Item & Setting \\
\midrule
EAF backbone & 335,987,842 parameters; 16 layers; dimension 768; 12 heads \\
Frozen WaveVAE & 10,823,361 parameters; 64-D latent; strides $[2,4,8,8]$ \\
Frozen duration predictor & 4,064,900 parameters \\
Optimizer & AdamW, $\epsilon=10^{-8}$, weight decay $10^{-2}$ \\
Base LR & $10^{-4}$; 5-epoch warmup; cosine decay to $10^{-5}$ \\
Condition LR & $2\times10^{-4}$ to $2\times10^{-5}$, applied only to frame/speaker conditioning groups \\
Flow-time sampling & bounded logit-normal on $(0.01,0.99)$: $t=0.01+0.98\,\mathrm{sigmoid}(u)$, $u\sim\mathcal N(-1.5,0.8^2)$ \\
Endpoint-loss time weight & $[1-\min(t,0.95)+10^{-5}]^{-2}$ \\
Gradient / EMA & max norm 1.0; EMA decay 0.999 \\
Loss weights & Flow 1.0; mode 1 latent 1.0; mel 1.0; STFT 1.0 \\
Mode 1 input perturbation & active Flow coordinate; 5\% of frames; Gaussian noise scale 0.1 \\
Budget & 4 GPUs; local batch 6; global batch 24; 100 epochs; 50,300 updates \\
ZCA regularization & relative floor $10^{-5}\lambda_{\max}$; absolute floor $10^{-8}$; both inactive here \\
Inference & 20 calls: 19 Euler updates to $t_{\mathrm{end}}=0.95$; one mode 1 pass \\
\bottomrule
\end{tabularx}
\end{table}

The base Flow backbone uses the base LR; the condition LR applies only to the frame- and speaker-conditioning parameter groups. The bounded logit-normal schedule is shared by the Scalar and ZCA mode 0 systems. Mode 1 does not use this continuous-time sampler: it is trained at fixed $t=1$ on the perturbed near-clean inputs described in Section 3.1. The endpoint-loss time weight is capped at $t=0.95$, so samples with $t>0.95$ receive a weight of approximately $400$ rather than the divergent $(1-t)^{-2}$ value. This mirrors the inference-side choice to stop mode 0 integration at $t_{\mathrm{end}}=0.95$. In the four-GPU runs, \texttt{DistributedSampler(drop\_last=True)} assigns 3,020 examples to each rank; the local loader then retains 503 complete batches of six and drops two examples per rank. Each epoch therefore contains 503 synchronized global batches ($12{,}072$ examples), drops eight examples in total, and yields $50{,}300$ updates over 100 epochs.

The research artifacts prepared for release include configuration files, scripts, covariance statistics, evaluation summaries, and environment information. We do not use RTF from the earlier historical runs as a primary comparison because accelerator metadata and a unified timing protocol were not preserved consistently. The matched sampling-efficiency sweep below instead reports RTF recorded by the same evaluation path for all cells; it remains an internal strict-GT generation-path measurement rather than an independently standardized latency benchmark.

\subsection{Inference-time sampling-efficiency audit}

We run a zero-retraining step sweep on Val399 strict-GT for all three matched raw-checkpoint pairs. The Euler endpoint remains $t_{\mathrm{end}}=0.95$, the solver remains Euler, and every cell retains one terminal mode 1 call. Total calls are 20, 10, 5, 3, and 2, corresponding to 19, 9, 4, 2, and 1 mode 0 updates, respectively. For $K$ total calls, the mode 0 step size is $\Delta t=0.95/(K-1)$, so every cell reaches the same mode 0 endpoint before refinement. No checkpoint is retrained or fine-tuned for a particular call count. The two-call cell is a large-step Euler diagnostic, not a MeanFlow or interval-average-velocity model.

The timed block begins at the strict-GT \texttt{synthesize\_alignment} call. It includes loading and preparing the cached alignment payload, phone-conditioning computation, all Flow calls, coordinate restoration, and frozen WaveVAE decoding. In ZCA cells, the inverse $64\times64$ transform is executed inside the timed decoder path and is therefore included; one-time checkpoint and covariance-statistics loading is outside the timer. The measurement excludes text normalization/G2P, duration prediction, waveform writing, ASR, and UTMOS evaluation. We therefore call it strict-GT generation-path RTF, use it only for matched internal quality--latency comparisons, and do not compare it directly with published end-to-end latency from other systems.

\subsection{Metrics and uncertainty}

Content is evaluated with \texttt{openai/whisper-large-v3} \cite{radford2022whisper} using corpus WER, mean utterance WER, P90 WER (the 90th percentile of utterance-level WER), and exact match after the same text normalization used for WER. Proxy naturalness is evaluated with UTMOS22 Strong (UTMOS hereafter), implemented by \texttt{tarepan/SpeechMOS:v1.2.0/utmos22\_strong} \cite{saeki2022utmos}, after resampling to 16 kHz. UTMOS22 was developed for VoiceMOS Challenge data derived mainly from Blizzard Challenge and Voice Conversion Challenge synthetic speech \cite{saeki2022utmos}; it was not calibrated to EAF-TTS outputs. Paired same-utterance differences are more defensible than absolute scores but remain proxies. We also report an internal DTW-MFCC distance, which is not directly comparable to public MCEP-based MCD values.

The primary full-step evaluation script computes normal-approximation intervals, $\bar x\pm1.96s/\sqrt n$, for system scores and paired utterance differences, where $n$ is the number of evaluated utterances (399 for Val399), not the number of training seeds. These intervals do not cluster by text length, phone composition, or speaker. The sampling-efficiency audit uses 20,000 paired utterance-bootstrap resamples. For WER and exact match, we additionally report a fixed-three-seed bootstrap that jointly resamples the same utterance IDs across all three trained systems; this quantifies uncertainty over the evaluated texts for these fixed seeds, not over unseen training seeds. The no-update gradient audit uses a separate paired batch bootstrap with 20,000 resamples. These bootstrap protocols are not pooled. Across training seeds, we report mean and seed SD only; three seeds are insufficient for a high-powered seed-level significance test. This is a retrospective controlled analysis, and the UTMOS endpoint was not publicly preregistered. WER, exact match, and DTW-MFCC are retained to reduce selective positive reporting. Future blind-text results should use a prospectively frozen paired analysis or a mixed-effects model.

\section{Results}

\subsection{The observed gap lies mainly in latent generation}

\begin{table}[H]
\centering
\caption{Test399 diagnostic comparison locating the observed proxy-naturalness gap. Test399 is not treated as blind.}
\label{tab:diagnostic-gap}
\small
\setlength{\tabcolsep}{4pt}
\begin{tabularx}{\textwidth}{@{}Yrrrr@{}}
\toprule
Test399 diagnostic system & WER & Mean WER & Exact & UTMOS \\
\midrule
Original waveform & 2.885\% & 3.237\% & 74.94\% & 4.372 \\
WaveVAE reconstruction & \textbf{2.856\%} & \textbf{3.176\%} & \textbf{75.44\%} & 4.259 \\
Scalar Flow, strict-GT & 3.162\% & 3.498\% & 71.93\% & 2.785 \\
\bottomrule
\end{tabularx}
\end{table}

WaveVAE reconstruction reduces UTMOS by only 0.114 and slightly reduces WER (2.885\% to 2.856\%). Scalar Flow reduces UTMOS by 1.474 relative to the reconstruction. This supports localizing the main observed proxy-naturalness gap to conditional latent generation. Test399 was repeatedly accessed during development; the table is diagnostic, not blind-test evidence.

\subsection{Val399 strict-GT results for fixed E100 raw checkpoints}

The following table reports the strict-GT path for three matched training seeds.

\begin{table}[H]
\centering
\caption{Val399 strict-GT results for fixed E100 raw checkpoints across three matched seeds.}
\label{tab:three-seed}
\scriptsize
\setlength{\tabcolsep}{3pt}
\begin{tabularx}{\textwidth}{@{}lYrrrrr@{}}
\toprule
Seed & System & WER & Exact & DTW-MFCC & UTMOS & ZCA\ensuremath{-}Scalar UTMOS \\
\midrule
42 & Scalar & \textbf{3.703\%} & \textbf{72.18\%} & \textbf{369.637} & 2.822 & --- \\
42 & ZCA & 3.896\% & 68.67\% & 370.720 & \textbf{3.199} & +0.377 \\
2027 & Scalar & 3.762\% & 71.43\% & 371.220 & 2.798 & --- \\
2027 & ZCA & \textbf{3.717\%} & \textbf{72.43\%} & \textbf{371.141} & \textbf{3.230} & +0.432 \\
3407 & Scalar & \textbf{3.717\%} & \textbf{71.68\%} & \textbf{371.706} & 2.844 & --- \\
3407 & ZCA & 3.851\% & 70.18\% & 372.002 & \textbf{3.249} & +0.405 \\
3-seed mean & Scalar & 3.727\% & 71.76\% & 370.854 & $2.822\pm0.023$ & --- \\
3-seed mean & ZCA & 3.822\% & 70.43\% & 371.288 & $3.226\pm0.025$ & $+0.405\pm0.027$ \\
\bottomrule
\end{tabularx}
\end{table}

All three seed-level UTMOS effects are positive, ranging from $+0.377$ to $+0.432$. For seed 3407, the paired difference is $+0.40507$ with normal-approximation 95\% CI $[+0.37882,+0.43132]$ and wins/ties/losses of 375/0/24. For seed 2027, the strict-GT raw-to-raw gain is $+0.432$; the EMA-to-raw comparison on the same strict-GT path (ZCA raw minus Scalar EMA under the existing selection rule) is $+0.44068$, with 95\% CI $[+0.41630,+0.46505]$ and 380/0/19 wins/ties/losses. These strict-GT effects are distinct from the text-only selected-checkpoint comparison in Section 5.4.

\begin{figure}[H]
\centering
\includegraphics[width=\textwidth]{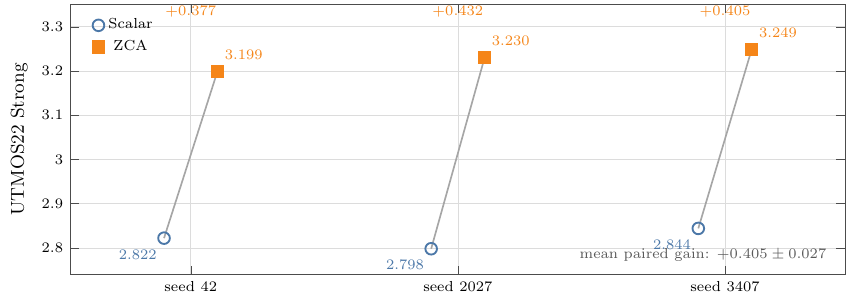}
\caption{Val399 strict-GT E100 raw UTMOS for three matched training seeds. Lines denote within-seed controlled comparisons. The summary is mean $\pm$ seed SD, not an utterance-level confidence interval.}
\end{figure}

Other metrics do not improve consistently. Averaged across seeds, ZCA increases WER by 0.095 percentage points, reduces exact match by 1.34 points, and increases DTW-MFCC by 0.43. Seed 2027 is the only seed for which ZCA is comparable or slightly better on all three content and reference-distance metrics. We do not have a mechanistic explanation for this seed-level variation and do not use it to support a general content claim. The across-seed consistency claim is therefore restricted to UTMOS proxy naturalness.

\subsection{Few-call inference without retraining}

\begin{figure}[H]
\centering
\includegraphics[width=\textwidth]{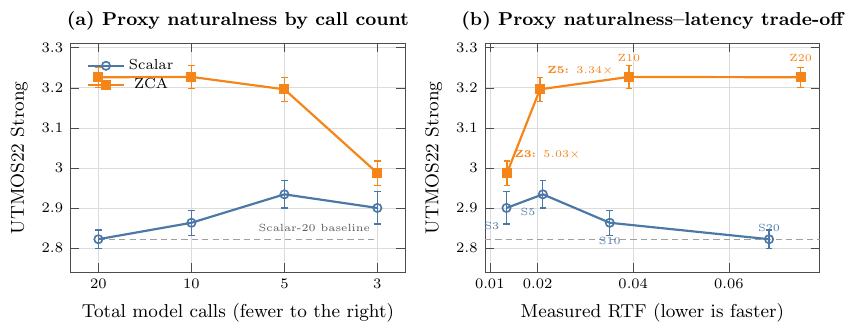}
\caption{Three-seed Val399 strict-GT sampling comparison. Left: UTMOS22 Strong versus total model calls. Right: the same proxy-naturalness values versus measured strict-GT generation-path RTF; labels identify Scalar (S) and ZCA (Z) call counts. Error bars denote seed SD. ZCA-5 and ZCA-3 speedups are measured against Scalar-20 and jointly change coordinates and call count. Lines connect inference configurations rather than training trajectories. The two-call diagnostic is retained in Table~\ref{tab:sampling-efficiency} but omitted here because both systems exhibit a large UTMOS collapse.}
\label{fig:sampling-tradeoff}
\end{figure}

Table~\ref{tab:sampling-efficiency} reports the matched inference-only sweep, and Figure~\ref{fig:sampling-tradeoff} visualizes the 20-, 10-, 5-, and 3-call quality--latency trade-off. UTMOS is mean $\pm$ seed SD; the remaining columns are three-seed means. Scalar-20 is the original baseline operating point used in the primary analysis.

\begin{table}[H]
\centering
\caption{Val399 strict-GT quality--latency sweep. Every cell uses the same endpoint and one terminal mode 1 call. RTF covers the strict-GT generation path defined in Section 4.4, including coordinate restoration and WaveVAE decoding.}
\label{tab:sampling-efficiency}
\scriptsize
\setlength{\tabcolsep}{3pt}
\begin{tabularx}{\textwidth}{@{}Yrrrrrr@{}}
\toprule
Coordinates & Calls & UTMOS & WER & Exact & DTW-MFCC & RTF \\
\midrule
Scalar & 20 & $2.822\pm0.023$ & 3.727\% & 71.76\% & 370.854 & 0.0684 \\
ZCA & 20 & $3.226\pm0.025$ & 3.822\% & 70.43\% & 371.288 & 0.0750 \\
Scalar & 10 & $2.863\pm0.031$ & 3.703\% & 72.18\% & 365.800 & 0.0351 \\
ZCA & 10 & $3.227\pm0.028$ & 3.846\% & 70.09\% & 367.177 & 0.0391 \\
Scalar & 5 & $2.934\pm0.034$ & 3.628\% & 72.60\% & 355.815 & 0.0211 \\
ZCA & 5 & $3.196\pm0.030$ & 3.668\% & 71.76\% & 359.225 & 0.0205 \\
Scalar & 3 & $2.900\pm0.040$ & 3.554\% & 73.18\% & 348.942 & 0.0135 \\
ZCA & 3 & $2.987\pm0.030$ & 3.584\% & 72.77\% & 351.964 & 0.0136 \\
Scalar & 2 & $2.348\pm0.032$ & 3.643\% & 72.01\% & 359.420 & 0.0100 \\
ZCA & 2 & $2.143\pm0.007$ & 3.727\% & 71.18\% & 366.498 & 0.0103 \\
\bottomrule
\end{tabularx}
\end{table}

ZCA-5 provides the strongest balanced deployment point. Relative to Scalar-20, it retains a $+0.375\pm0.027$ UTMOS advantage, lowers DTW-MFCC by 11.629, and reduces measured RTF from 0.0684 to 0.0205 ($3.34\times$ using the rounded table values). In the fixed-three-seed paired utterance bootstrap, its corpus-WER difference is $-0.059$ percentage points (95\% CI $[-0.202,+0.079]$), while the exact-match difference is 0.000 points (95\% CI $[-1.337,+1.337]$). Thus, the evaluated utterances provide no evidence that this quality--latency point sacrifices ASR-derived content accuracy.

ZCA-3 is a more aggressive point: it remains $+0.165\pm0.031$ UTMOS above Scalar-20 while reducing measured RTF to 0.0136 ($5.03\times$ using rounded values). Its fixed-three-seed WER difference is $-0.144$ percentage points (95\% CI $[-0.286,-0.010]$); its exact-match difference is $+1.003$ points, although the interval crosses zero ($[-0.418,+2.423]$). The WER reduction is driven mainly by 0.119 fewer substitutions per 100 reference words rather than increased deletions or insertions. These intervals quantify evaluated-text uncertainty for the three fixed training seeds and do not provide high-powered inference over unseen training runs.

Reducing calls benefits both coordinate systems, so this result is not evidence that ZCA intrinsically improves low-step integration. Define the UTMOS robustness interaction at $K$ calls as

\[
\bigl(\mathrm{ZCA}_K-\mathrm{ZCA}_{20}\bigr)
-\bigl(\mathrm{Scalar}_K-\mathrm{Scalar}_{20}\bigr).
\]

The interaction is negative for every reduced-call condition and all three seeds; every per-seed paired 95\% CI excludes zero. Across seeds, the point estimates range from $-0.045$ to $-0.036$ at 10 calls, $-0.148$ to $-0.136$ at five calls, $-0.330$ to $-0.299$ at three calls, and $-0.633$ to $-0.589$ at two calls. ZCA therefore loses its relative advantage faster than Scalar as calls decrease. Its practical value is instead that the higher full-step UTMOS score supplies enough headroom for ZCA-5 and ZCA-3 to remain above Scalar-20 while using fewer calls. At two calls, ZCA UTMOS collapses to 2.143 even though WER remains 3.727\%; this divergence shows that ASR content metrics alone cannot detect the acoustic-naturalness failure.

\subsection{Text-only selected-checkpoint deployment comparison}

The seed-2027 text-only Val399 comparison reads no MFA alignment, phone cache, or duration cache. It compares Scalar E100 EMA with ZCA E100 raw selected under the existing, non-preregistered validation rule.

\begin{table}[H]
\centering
\caption{Seed-2027 selected-checkpoint text-only deployment comparison on Val399.}
\label{tab:text-only}
\scriptsize
\setlength{\tabcolsep}{3pt}
\begin{tabularx}{\textwidth}{@{}Yrrrrrrr@{}}
\toprule
System & WER & \shortstack{Mean\\WER} & \shortstack{P90\\WER} & Exact & DTW-MFCC & UTMOS & 95\% CI \\
\midrule
Scalar E100 EMA & \textbf{5.175\%} & \textbf{5.388\%} & 18.355\% & \textbf{62.66\%} & \textbf{404.680} & 2.622 & [2.596, 2.648] \\
ZCA E100 raw & 5.234\% & 5.513\% & \textbf{17.271\%} & 60.65\% & 409.332 & \textbf{3.053} & [3.023, 3.083] \\
\bottomrule
\end{tabularx}
\end{table}

\emph{The system-level intervals in the final column are not the paired-difference interval; the paired comparison is reported below.}

The paired UTMOS difference is $+0.43111$ with normal-approximation 95\% CI $[+0.40400,+0.45821]$ and 374/0/25 wins/ties/losses on Val399. The selected deployment system therefore retains the proxy-naturalness gain without test-time acoustic caches. Because the weight types differ, the full difference cannot be attributed to ZCA. Matched raw/raw and EMA/EMA text-only comparisons remain required.

Using the unrounded values, corpus WER increases by 0.0595 percentage points (5.175\% to 5.234\% after rounding), mean WER by 0.1244 percentage points, exact match decreases by 2.01 percentage points, and DTW-MFCC increases by 4.652. P90 WER improves by 1.084 percentage points, and neither system produces a pathological sample. The result is a proxy-naturalness/content trade-off, not a uniform quality improvement.

\subsection{Test399 diagnostics (non-blind)}

\textbf{Non-blind Test399 diagnostics for the seed-42 fixed E100 non-EMA checkpoints across three inference paths.}

\begin{table}[H]
\centering
\caption{Non-blind Test399 diagnostics for the seed-42 fixed E100 raw checkpoints.}
\label{tab:test399-paths}
\small
\setlength{\tabcolsep}{3pt}
\begin{tabularx}{\textwidth}{@{}Yrrrrr@{}}
\toprule
Condition & Scalar WER & ZCA WER & Scalar UTMOS & ZCA UTMOS & UTMOS gain \\
\midrule
Strict-GT & \textbf{3.162\%} & 3.293\% & 2.785 & \textbf{3.163} & +0.378 \\
Cached predicted duration & \textbf{4.022\%} & 4.677\% & 2.582 & \textbf{2.996} & +0.414 \\
Text-only & \textbf{4.240\%} & 4.400\% & 2.575 & \textbf{2.992} & +0.417 \\
\bottomrule
\end{tabularx}
\end{table}

The proxy-naturalness gain appears across all three paths. Test399 text-only has 375/399 ZCA wins, whereas the separate Val399 text-only comparison in Section 5.4 has 374/399. These comparisons use different splits and checkpoints, so the near-identical win counts are coincidental rather than a replication. The larger WER cost under predicted durations prevents interpreting the result as cost-free deployment improvement.

\section{Analysis and Discussion}

\subsection{What the results support}

Scalar Flow maps a unit Gaussian to a target with covariance condition number 1355.44. ZCA equalizes second-order principal-direction scales and may reduce source--target covariance mismatch or high-variance domination of Euclidean regression. The three matched-seed effects and text-only deployment result are consistent with improved conditional distribution modeling rather than improved codec reconstruction alone.

The evidence does not isolate condition number as the cause. Full ZCA rotates and rescales coordinates and changes both Flow and auxiliary-loss metrics. We have not yet completed LJSpeech partial-whitening, rotation-only, or reference-coordinate-loss interventions. The supported claim is the aggregate effect of full ZCA with train-data-estimated statistics.

\subsection{Naturalness headroom can be exchanged for fewer calls}

The sampling sweep adds a practical but distinct finding. It does not show that ZCA improves the numerical order, stability region, or low-step accuracy of Euler integration: every robustness interaction is negative. Scalar itself improves at moderate call reductions, which suggests that the 20-call rollout may accumulate local endpoint-prediction errors. Instead, full-step ZCA begins from a substantially higher UTMOS level. Reducing its calls spends part of this margin while improving RTF and, at three to five calls, ASR-derived content and DTW-MFCC point estimates.

This distinction matters for deployment claims. The supported statement is that \emph{ZCA supplies proxy-naturalness headroom that can be exchanged for measured inference speed under this evaluator}. It is not evidence that whitening intrinsically accelerates ODE solving. ZCA-5 is the balanced operating point; ZCA-3 is the aggressive low-latency point; the two-call result is a failure boundary. Because the RTF excludes text normalization/G2P and duration prediction and was not obtained from an independent latency harness, we do not compare it with published end-to-end RTF values.

\subsection{Covariance whitening is not perceptual whitening}

We estimate frozen-decoder log-mel sensitivity on 32 Val utterances, four boundary-safe five-frame windows each (128 windows). Let $q_i$ and $\lambda_i$ denote covariance eigenvectors and eigenvalues, and define

\[
s_i=\mathbb E_z\left[\left\|J_{\operatorname{logmel}\circ D}(z)q_i\right\|_2^2\right].
\]

A unit direction in ZCA coordinates maps through the inverse transform to $\sqrt{\lambda_i}q_i$, so its diagonal sensitivity proxy is $\lambda_i s_i$. We report

\[
\kappa_{\mathrm{diag}}^{\mathrm{raw}}=\frac{\max_i s_i}{\min_i s_i},\qquad
\kappa_{\mathrm{diag}}^{\mathrm{ZCA}}=\frac{\max_i\lambda_i s_i}{\min_i\lambda_i s_i}.
\]

Spearman correlation between $\lambda_i$ and $s_i$ is 0.276. The raw diagonal proxy is 30.54, the ZCA proxy is 21,740.76, and the ZCA P95/P05 ratio is 125.84. The much larger ZCA proxy arises after multiplying $s_i$ by $\lambda_i$, whose dynamic range (1355.44) is far larger than that of $s_i$ (30.54); their weak rank correlation means that this covariance scaling need not equalize decoder sensitivity. In other words, equalizing data variance can preserve or amplify decoder-metric anisotropy when high-variance directions are also decoder-sensitive. These are local diagonal proxies in the covariance eigenbasis, not full spatiotemporal Gram condition numbers. A small finite-difference check uses a latent-coordinate perturbation of magnitude $\epsilon=0.1$ along unit covariance eigenvectors. It yields a finite-difference/JVP ratio median of 1.127, with 87.5\% of directions within $[0.5,2]$, indicating that JVP-based sensitivity agrees in scale with finite differences for most checked directions. This supports scale validation but not the extreme-value condition proxy.

ZCA therefore whitens data covariance but not the decoder-induced perceptual metric. The audit also does not explain the positive UTMOS result: proxy anisotropy becomes larger while generated UTMOS improves.

\subsection{Gradient conflict is not a simple harm indicator}

On 32 matched no-update batches, ZCA changes the latent--mel and mel--STFT gradient cosines by $-0.102$ and $-0.377$, respectively. The mean change across three decoder-related pairs is $-0.159$, with a 20,000-resample paired batch-bootstrap 95\% CI of $[-0.185,-0.129]$. This bootstrap is specific to the gradient audit; the primary full-step UTMOS intervals use the normal approximation described in Section 4.5. ZCA thus increases several local multi-objective conflicts even though final UTMOS improves substantially.

This negative audit shows that stronger local gradient conflict is not sufficient to predict worse generation. It may reflect a redistribution of responsibilities among latent, mel, and STFT objectives. We do not infer that gradient surgery or dynamic loss weighting would be ineffective.

\subsection{Proxy naturalness versus content}

At the primary 20-call setting, UTMOS improves for all three seeds, while WER, exact match, and DTW-MFCC are seed dependent. ZCA may increase the effective weight of low-variance acoustic directions that affect timbre, harmonics, or local dynamics preferred by the UTMOS predictor without improving phone boundaries or ASR confidence. The few-call deployment comparisons show that the ZCA UTMOS margin can be retained without detected paired content degradation relative to Scalar-20, but they do not establish that ZCA intrinsically improves intelligibility. Current results cannot distinguish human-perceived improvement from predictor preference.

We therefore treat ZCA as a candidate \textbf{UTMOS proxy-naturalness preconditioner}, not a universal replacement for Scalar coordinates. As an illustrative post-hoc deployment gate, suppose WER may worsen by at most 0.05 percentage points, exact match by at most 0.5 points, and pathological count may not increase. The selected seed-2027 text-only result fails this gate (WER $+0.0595$, exact $-2.01$). This illustrative gate was not preregistered and was not used to select any reported system; deployment requires a prospectively frozen gate, matched checkpoint controls, and listening tests.

\subsection{Engineering properties}

ZCA statistics are computed once offline, add no trainable parameters, and require only a 64-dimensional linear inverse before decoding. The interface is independent of text and test audio and can be used in strict-GT, predicted-duration, and text-only paths. It does require retraining the Flow: a Scalar-trained checkpoint cannot be switched to ZCA coordinates post hoc. Once trained, however, the same checkpoint can be evaluated at the three- and five-call operating points without additional optimization.

\section{Limitations and Threats to Validity}

\textbf{Scope and external validity.} UTMOS22 Strong is an automatic naturalness proxy, not human MOS. It was not calibrated to EAF-TTS outputs or publicly preregistered as the primary endpoint, and no human listening study was conducted. Absolute proxy scores remain below the frozen-codec reconstruction reference, so the reported gain should not be read as approaching human naturalness. The study covers one single-speaker dataset, one WaveVAE checkpoint, one latent dimension, and one EAF-TTS configuration. We make no cross-speaker, cross-dataset, cross-codec, multilingual, noisy, expressive, or style-controlled generalization claim, and we do not compare EAF-TTS end to end with external TTS systems.

\textbf{Internal validity and causal attribution.} Full ZCA jointly changes the training and sampling coordinates, the induced codec-space loss metric, and mode 1 perturbation geometry. The reported result is therefore a full-protocol effect, not a causal effect of lowering the covariance condition number alone. We do not report partial whitening, rotation-only, per-channel scaling, reference-coordinate loss, inference-only ZCA, or conditional/local whitening controls. Padding is excluded by the valid-frame mask, but sensitivity to additionally excluding silence or short pauses is not reported. Mode 1 is trained on perturbed clean latents but receives mode 0 ODE outputs at inference, so its training and inference input distributions are not assumed to match. Its numeric perturbation scale is matched in each active Flow coordinate rather than in codec space. We did not run codec-space-matched perturbation, no-mode-1, or mode-0-output-fed mode 1 ablations. The text-only selected-checkpoint comparison mixes Scalar EMA and ZCA raw checkpoints; it is confounded by checkpoint type and is treated as directional deployment evidence, not as a causal coordinate estimate.

\textbf{Measurement and statistics.} Test399 was repeatedly accessed and is not blind-test evidence. The primary evidence uses text-disjoint Val399, but Val399 is a development validation set rather than a prospectively untouched blind test set. Scalar and ZCA deliberately share hyperparameters, schedule, and checkpoint rule to isolate the coordinate intervention; neither was tuned separately, so the study does not establish optimality under per-system tuning. Three seeds are insufficient for high-powered seed-level inference. Primary full-step UTMOS intervals use normal approximations without clustering by text length, phone composition, or speaker; sampling and gradient audits use separate paired bootstrap protocols rather than one mixed-effects analysis. At 20 calls, WER, exact-match rate, and DTW-MFCC do not improve consistently, so the positive effect is localized mainly to proxy naturalness rather than content fidelity. Few-call paired content intervals quantify uncertainty over the evaluated texts for three fixed seeds, not over new training runs; intervals containing zero support ``no statistically detectable degradation,'' not equivalence. The internal DTW-MFCC distance is not directly comparable with public MCEP-based MCD. Sampling-sweep RTF includes Flow calls, coordinate restoration, and frozen WaveVAE decoding, but excludes text normalization/G2P, duration prediction, waveform writing, ASR, and UTMOS. It is not an independently standardized end-to-end latency benchmark and should not be compared directly with published end-to-end RTF.

\textbf{Mechanistic interpretation.} Frozen-decoder and gradient audits, together with inconsistent content/reference-distance results, constrain but do not establish a mechanism. We do not show that ZCA improves generation by lowering the condition number, or that frozen-decoder sensitivity aligns with data covariance. At every reduced call count, ZCA's advantage over Scalar is smaller than at 20 calls. The few-call results therefore demonstrate usable full-step proxy-naturalness headroom, not intrinsically better low-step integration. A mechanistic account, a prospectively untouched test set, and broader validation remain open.

\section{Conclusion}

LJSpeech WaveVAE training latents are strongly anisotropic even though the frozen codec reconstructs real latents well. A fixed ZCA interface with no learned parameters, whose statistics are estimated only from training frames, raises Val399 E100 raw UTMOS for all three matched training seeds, with mean gain $+0.405\pm0.027$. A selected-checkpoint text-only comparison also gains $+0.4311$ without test-time acoustic caches, although it mixes EMA and raw checkpoints. Frozen-decoder sensitivity and gradient audits constrain simple explanations.

The inference-only sweep shows that this proxy-naturalness margin can be exchanged for fewer model calls. Relative to Scalar-20, ZCA-5 retains a $+0.375\pm0.027$ UTMOS advantage with no statistically detectable paired WER or exact-match degradation for the three fixed seeds, although equivalence is not established, while reducing measured strict-GT generation-path RTF by $3.34\times$. ZCA-3 retains a smaller $+0.165\pm0.031$ advantage at a $5.03\times$ operating point, whereas two calls cause a severe UTMOS collapse. Because all reduced-call robustness interactions are negative, the result is not evidence that ZCA intrinsically improves low-step integration; it reflects usable naturalness headroom.

The evidence supports a narrow conclusion: \textbf{within this single-speaker, single-WaveVAE LJSpeech/EAF-TTS protocol, global ZCA with train-data-only statistics consistently improves UTMOS automatic proxy naturalness across three matched seeds and enables favorable measured few-call operating points.} It has not yet been shown to improve human perception or to provide a general speed advantage across hardware, datasets, codecs, or TTS architectures. Blind text, listening tests, matched text-only checkpoints, and coordinate-versus-metric ablations remain necessary.

\section{Data, Ethics, and Code Availability}

LJSpeech contains 13,100 English utterances from one speaker and is released in the public domain \cite{ito2017ljspeech}. This study uses public audio and existing automatic metrics and recruits no new human participants. Future listening tests will follow applicable ethics review, informed-consent, and data-retention requirements.

Scalar/ZCA configurations, covariance-statistics code, evaluation scripts, and result summaries are available from the authors upon request. A versioned public archive with checkpoint hashes and an environment lockfile is in preparation.

\section{Future Work}

\begin{enumerate}
\item Freeze 300--500 new blind texts and evaluate once; conduct randomized A/B or CMOS tests on 50--100 items for naturalness, clarity, and content errors.
\item Complete the text-only checkpoint matrix: Scalar raw/ZCA raw and Scalar EMA/ZCA EMA.
\item Separate coordinate and metric effects with per-channel scaling, rotation-only, partial whitening, and reference-coordinate auxiliary loss.
\item Repeat the few-call comparison on prospectively frozen blind text, use a mixed-effects analysis across additional training seeds, and stratify content errors by text length and phone composition.
\item Add a second WaveVAE checkpoint and, if feasible, a second single-speaker corpus.
\end{enumerate}

\begingroup
\small
\setlength{\bibsep}{2pt}
\bibliographystyle{unsrtnat}
\bibliography{references}
\endgroup

\end{document}